\documentstyle[preprint,aps,epsf]{revtex}

\begin{document}
\preprint{hep-lat/9604003}
\title{
Random matrix model of QCD at finite density and
the nature of the quenched limit
}
\author{
M. A. Stephanov
\thanks{Electronic mail address: \it misha@uiuc.edu.}}
\address{
	Department of Physics, 
	University of Illinois at Urbana-Champaign,\\
	1110 West Green Street, Urbana, IL 61801-3080, USA}
\date{March 1996}
\maketitle
\begin{abstract}
We use a random matrix model to study chiral symmetry breaking in QCD
at finite chemical potential $\mu$. We solve the model and compute the
eigenvalue density of the Dirac matrix on a complex plane. A naive
``replica trick'' fails for $\mu\neq0$: we find that quenched QCD is
not a simple $n\to0$ limit of QCD with $n$ quarks. It is the limit of
a theory with $2n$ quarks: $n$ quarks with original action and $n$
quarks with conjugate action. The results agree with earlier studies
of lattice QCD at $\mu\neq0$ and provide a simple analytical
explanation of a long-standing puzzle.
\end{abstract}

\newpage
\narrowtext

\section{Introduction}

The spontaneous breaking of chiral symmetry is one of the most
important dynamical properties of QCD which shapes the hadronic
spectrum. A great deal of understanding of this nonperturbative
phenomenon at zero and finite temperature has been achieved by various
methods \cite{Sh95}. In particular, we expect that the chiral symmetry
is restored above a certain critical temperature. The study of this
new chirally symmetric phase of hot QCD is one of the primary
objectives of heavy ion colliders. In contrast, the behavior of QCD at
large baryon density (conditions which can arise in the heavy ion
colliders or in neutron stars) is not well understood. The main puzzle
has for a long time been a contradiction between a straightforward
physical expectation and numerical results from quenched lattice
QCD~\cite{Ba86,KoLo95}. Simulations with dynamical quarks, on the
other hand, are very inefficient at finite $\mu$ --- the fermion
determinant is complex.

The puzzle concerns the dependence of the order parameter (the chiral
condensate $\langle\bar\psi\psi\rangle$) on the baryon chemical
potential. A non-analytical change in the value of
$\langle\bar\psi\psi\rangle$ should occur when $\mu > \mu_c \approx
m_B/3$, where $m_B$ is the mass of the lightest baryon. At this point
the production of baryons becomes energetically favorable. For smaller
$\mu$ the value of $\langle\bar\psi\psi\rangle$ is nonzero. In
contrast, lattice simulations of quenched QCD indicate that $\mu_c=0$
(at zero bare quark mass), i.e., the chiral condensate vanishes if
$\mu\neq0$ \cite{Ba86,KoLo95}. A number of possible explanations has
been suggested \cite{explain}. However, the answer to this puzzle
remains unclear.

This work was motivated by a desire to shed some light on this
question using the random matrix approach which received considerable
interest recently
\cite{ShVe93,VeZa93,Ve94,JaVe95,WeSc96,St96,JaSe96,JuNo96}. It is
based on the idea that, for the purpose of studying chiral symmetry
breaking, fluctuations of the Dirac operator in the background of the
gauge fields can be approximated by purely random fluctuations of its
matrix elements in a suitable basis. For example, in the instanton
liquid model this basis can be formed from the Dirac zero modes for
individual (anti)instantons, which due to overlaps form a band of
small eigenvalues responsible for the chiral symmetry
breaking~\cite{DiPe86}. A similar random matrix approach is 
fruitful in the studies of spectra of systems with a high level of
disorder, such as spectra of heavy nuclei \cite{Po65}. Introduction of
chemical potential into such a model of chiral symmetry breaking is
straightforward. The resulting Dirac matrix (times $i$) is {\em
non-hermitian}. Thus the eigenvalues lie in the complex plane rather
than on a line. Such random matrix models have not received much
attention previously and this study is a step in an unexplored
direction.

In this Letter we show how to solve such a model in the
thermodynamic limit and discuss the implications.

\section{The resolvent}

In order to study chiral symmetry breaking we shall calculate 
the resolvent of the Dirac operator $D$:
\begin{equation}\label{G}
G = \left\langle \mbox{ tr }(z - D)^{-1} \right\rangle,
\end{equation}
as a function of the bare quark mass $z$ which we take to be a
complex variable $z=x+iy$. The average is over fluctuations of the
random matrix elements of $D$. It should be obvious that $G$ is the
same as $\langle\bar\psi\psi\rangle$.
The resolvent can be expressed through the average
eigenvalue density $\rho$:
\begin{equation}\label{Grho}
G(x,y)=\int\,dx'\,dy'\,\rho(x',y'){1\over z - z'}.
\end{equation}
A vector $\vec G =(\mbox{Re}G, -\mbox{Im}G)$ is the
electric field created by the charge distribution $\rho$.
This makes the inversion of (\ref{Grho}) obvious:
\begin{equation}\label{rhoG}
\rho={1\over2\pi}\vec\nabla\vec G 
= {1\over\pi}{ \partial \over \partial z^* } G,
\end{equation}
where $\partial/\partial z^* 
\equiv (\partial/\partial x +i \partial/\partial y)/2$.

Analytical properties of $G$ are very closely related to the chiral
symmetry breaking. 
From (\ref{rhoG}) we see that $\rho$ vanishes if the function $G$ is
holomorphic. A discontinuity of $G$ along a cut going through $z=0$ is
the signature of the spontaneous chiral symmetry breaking:
$\langle\bar\psi\psi\rangle(+0)\neq\langle\bar\psi\psi\rangle(-0)$.
This observation together with~(\ref{rhoG}) leads to the
Banks-Casher relation \cite{BaCa80}: $\langle\bar\psi\psi\rangle = \pi
\rho(0)$, where $\rho(0)$ is the density per {\it length} on the cut at
$z=0$. However, (\ref{rhoG}) is more general and can be applied to a
case when the non-analyticity is not in the form of a cut but occupies
a 2-dimensional patch, which is the case in our model.

\section{The matrix model and naive replica trick}

The matrix $D$ has the form:
\begin{equation}
D=\left(
\begin{array}{cc}
0         & iX\\
iX^\dagger & 0
\end{array}
\right) + \left(
\begin{array}{cc}
0   & \mu \\
\mu & 0
\end{array}
\right)
\end{equation}
where we added the chemical potential term $\mu\gamma_0$ to the Dirac
matrix \cite{ShVe93}. The $N\times N$ matrix elements of $X$ are
independently distributed complex Gaussian random variables: $P(X) =
\mbox{const}\times\exp\{-N\mbox{ Tr }XX^\dagger\}$. The unit of mass
in the model is set by $n_4/\langle\bar\psi\psi\rangle_0 \sim
200\mbox{ MeV}$, where $n_4\sim1\mbox{ fm}^{-4}$ is the number of small
eigenvalues in a unit volume (instanton density~\cite{Sh88}) and
$\langle\bar\psi\psi\rangle_0\sim(200\mbox{ MeV})^3$ is the chiral
condensate at $T=0$, $\mu=0$.

In order to find the resolvent (\ref{G}) we introduce ${n}$ 
quark fields (replicas) and calculate:
\begin{equation}\label{Vn}
V_{{n}}=-{1\over{n}}\ln \left\langle {\det}^{{n}} ( z - D ) \right\rangle.
\end{equation}
This quantity continued to ${n}\to0$ (quenched limit) becomes:
\begin{equation}
V = -\left\langle \ln \det ( z - D ) \right\rangle,
\end{equation}
from which we find $G$:
\begin{equation}\label{GV}
G = -{\partial\over\partial z} V.
\end{equation}
The trace in (\ref{G}) is normalized as $\mbox{ tr }1=\mbox{Tr }1/(2N)$.
Following the electrostatic analogy of the previous section
one can view Re$V$ as the scalar potential for $\vec G$.

Using Hubbard-Stratonovitch transformation we obtain:
\begin{equation}\label{expV}
\exp\{-{n} V_{{n}}\}
= \int {\cal D} a\, {\det}^N \left(
\begin{array}{cc}
z+a    &  \mu         \\
\mu    &   z+a^\dagger
\end{array} \right)
\exp\{-N\mbox{ Tr }aa^\dagger\},
\end{equation}
where $a$ is an auxiliary complex ${n}\times {n}$ matrix field.  For
large $N$ the calculation of the integral amounts to finding its
saddle point.  If we assume that the replica symmetry is not broken
(i.e., $a$ is proportional to a unit matrix) we arrive at the saddle
point equation:
\begin{equation}\label{a}
(z+a)=a[(z+a)^2-\mu^2],
\end{equation}
The complex value of $a$ in (\ref{a}) is the analytical continuation
of the real part of the diagonal matrix elements of $a$ in (8).  The
imaginary part is zero in the saddle point.
The solution of this cubic equation is
straightforward. It is the same as in a similar model \cite{St96}
with $\omega\to i\mu$.
Finally, it is easy to find using (\ref{GV},\ref{expV}) that $G=a$
where $a$ is the saddle point given by (\ref{a}).  

The $V_{n}$ does not depend on ${n}$ and the limit ${n}\to0$ seems
obvious.  However, in the next section we shall compare this
expectation to numerical data and see that the limit $n\to0$ is in
fact very different! Now let us summarize the properties of this model
for $n>0$.

We see that $G(z)$ is a holomorphic function.  It has 3 Riemann sheets
and we select the one where $G\to1/z$ for $z\to\infty$ -- which
follows from $\int dxdy\,\rho=1$ and the Gauss theorem. The only
singularities on the physical sheet are the pole at $z=\infty$ and 2
(for $\mu^2\le1/8$) or 4 (for $\mu^2>1/8$) branch points connected by
cuts. The brunch points are where 2 of the 3 solutions of (\ref{a})
coincide. The trajectory of a cut is determined by a condition that
$a$ is the deepest minimum (out of 3) of: $\mbox{ Re } [a^2 -
\ln((z+a)^2-\mu^2)]$.

At $\mu=0$ the cut along imaginary axis connects two singularities at
$z=\pm2i$. For nonzero $\mu$ the singularities start moving towards
each other along the imaginary axis. At $\mu^2=1/8$ each of the branch
points bifurcates in two ones which move off the $y$ axis into the
complex plane. The cut goes through the origin (along the $y$ axis)
until $\mu^2=0.278...$. At this point it splits into two cuts
connecting complex conjugate points.  This means that
$\mu_c^2=0.278...$ in such a model.

\section{Numerical results and the solution of the model}

For $n=0$ one can easily determine the density of eigenvalues
numerically by calculating the eigenvalues of the random matrix $D$
and plotting them on a complex plane. The density of points on such a
scatter plot is proportional to $\rho$. The results for different
values of $\mu^2$ are shown in Fig. 1. They contradict naive
expectations from the previous section. At $\mu=0$ all eigenvalues are
distributed between points $z=\pm2i$ on the $y$ axis. However, already
at very small nonzero $\mu^2\ll1/8$ the eigenvalue density is nonzero
in a ``blob'' of finite width in $x$ direction which grows with
$\mu$. The same behavior is seen in quenched lattice QCD~\cite{Ba86}
and gives rise to the paradox described in the Introduction: there is
no discontinuity in the value of $\langle\bar\psi\psi\rangle$ at any
$\mu>0$. The matrix model has an advantage: it is amenable to exact
treatment which clarifies the nature of the problem.

The failure of the naive replica approach can be understood if we look
at the expression~(\ref{Vn}): it does not contain $z^*$!  On the other
hand, eq.~(\ref{rhoG}) tells us that $\rho\neq0$ if $G$ depends on
$z^*$, i.e., if it is not holomorphic. In fact, the correct replica
trick for a non-hermitian matrix should start from the quantity:
\begin{equation}\label{Vn2}
V_{{n,n}} = -{1\over {n}}\ln\left\langle 
{\det}^{{n}} (z - D) (z^* - D^\dagger) \right\rangle,
\end{equation}
which is now real due to introduction of the quarks with conjugate
Dirac matrix.
Naively, in the limit ${n}\to0$ the conjugate quarks decouple but, as
we shall see, this is not always the case!
In mathematics an analogous construction is called a
V-transform~\cite{Gi88} and allows one to study spectra of
non-hermitian matrices. In the present context this formal
construction has a clear and simple physical meaning.

We can calculate (\ref{Vn2}) using the same method as for (\ref{Vn}).
Now, however, we have to introduce 4 auxiliary complex ${n}\times {n}$
fields, and we arrive at:
\begin{eqnarray}
\exp(- {n} V_{{n,n}})
&=& \int {\cal D}a\,{\cal D}b\,{\cal D}c\,{\cal D}d\,\,
{\det}^{N} \left(
\begin{array}{cccc}
z+a  		& \mu      	&   0      		&  id 	\\
\mu  		& z+a^\dagger   &   ic   		&  0  	\\
0    		& id^\dagger    &   z^*+b^\dagger	&  \mu 	\\
ic^\dagger   	& 0         	&   \mu    		&  z^*+b 
\end{array}
\right) 
\nonumber \\&&
\times\exp\left\{-N(|a|^2 + |b|^2 + |c|^2 + |d|^2)\right\}.
\end{eqnarray}

The set of solutions of the saddle point equation is richer in this
case. There is a solution with $c=d=0$. In this case the conjugate
quarks do decouple and we obtain the same holomorphic function $G$ as
before. However, there is another solution in which the condensates
$c$ and $d$ are not zero! Then the function $G$ is not holomorphic and
therefore $\rho\neq0$. This saddle point dominates the integral at
small $z$ for $0<\mu<1$.

The condensates $c$ and $d$ are bilinears of the type
$\langle\bar\psi\chi\rangle$, mixing original $\psi$ and conjugate
$\chi$ quarks. These condensates do not break the original chiral
symmetry but a spurious (replica type) symmetry involving both
original and conjugate quarks. Similar condensates carrying baryon
number were discussed in the $SU(2)$ model of QCD {\em with} quarks
\cite{Da86-7}.  In the quenched theory, as in \cite{Da86-7}, the
original chiral symmetry is always restored at $\mu>0$. The spurious
symmetry is spontaneously broken for $\mu<1$ and is restored for
$\mu>1$.

The boundary of the $\rho\neq0$ region is given by:
\begin{equation}\label{yx}
y^2 = (\mu^2 - x^2)^{-2} 
[4 \mu^4 (1 - \mu^2) - (1 + 4\mu^2 - 8\mu^4) x^2 - 4 \mu^2 x^4].
\end{equation}
It is plotted on Fig. 1 for comparison with numerical data.
The baryonic condensates $c$ and $d$ inside of the ``blob'' are
given by:
\begin{equation}
|c|^2=|d|^2=
{\mu^2\over\mu^2-x^2} - \mu^2 - {x^2\over4(\mu^2-x^2)^2} 
- {y^2\over4}.
\end{equation}
On the boundary (\ref{yx}) they vanish and the two solutions
(holomorphic and non-holomorphic) match. In the outer region:
$c=d=0$ and $G=a$ is the solution of the cubic equation
(\ref{a}).  Inside of the ``blob'' the resolvent is given by:
\begin{equation}
G = a = {1\over2} {x\over \mu^2 - x^2} - x - {iy\over 2},
\end{equation}
and the density of the eigenvalues (\ref{rhoG}) is:
\begin{equation}
\rho = {1\over4\pi}\left( {{x^2+\mu^2\over(\mu^2-x^2)^2} - 1}\right).
\end{equation}

To appreciate non-triviality of this result one should notice that
expression (\ref{G}) which defines the resolvent appears to
depend only on $z$! The limit $n\to0$ must be taken with great
care, as is well-known in the replica approach~\cite{MePa87}.

\section{Conclusions}

The fermion determinant in QCD is complex at nonzero chemical
potential. Lattice simulations of such a theory are extremely
inefficient. Therefore all reliable data from lattice QCD so far have
been obtained for a quenched theory.  We learn from the random matrix
model that the quenched theory at finite $\mu$ behaves
qualitatively different from the QCD with dynamical quarks. Rather,
the quenched approximation describes a theory where each of the quarks
has a conjugate partner, so that the fermion determinant is
non-negative. We see that for such a theory the result $\mu_c=0$ is
natural.  Similar arguments have been given by several authors in
different settings and using less realistic models
\cite{explain}. Here it can be demonstrated in a very clean and
explicit way.

Simulations with dynamical quarks at strong coupling are possible in
$SU(2)$ and $SU(4)$ QCD~\cite{Da86-7} and also agree with our
results. The $\mu_c$ is finite in the $SU(4)$ theory. On the other
hand, in the $SU(2)$ theory, where the quarks are self-conjugate,
$\mu_c=0$ due to the baryonic condensates.

The matrix model describes many features of the chiral symmetry
breaking in QCD very well
\cite{ShVe93,VeZa93,Ve94,JaVe95,WeSc96,St96,JaSe96,JuNo96}. One of the
apparent limitations, however, is that it is static --- there are no
kinetic terms and we cannot study spectrum of masses. In the quenched
QCD $\mu_c$ appears to coincide with half of the mass of the so-called
baryonic pion~\cite{explain} --- a bound state of a quark and a
conjugate antiquark. It is degenerate with the $\pi$-meson but carries
a nonzero baryon number. From the exact solution (\ref{yx}) we find
$\mu_c\approx\sqrt{m/2}$, for small quark mass $m\ll1$. If we had
$f_\pi$ in our model we could relate $\mu_c$ to the mass of the
pion. The model also does not account for the confinement of quarks.
It remains to be seen if the confinement plays a role in the case
under consideration.

\acknowledgments

The author is grateful to E. Fradkin, A. Kocic, J.~Kogut, and
M.~Tsypin for discussions and suggestions some of which were crucial
for this work.
The work was supported by the National Science Foundation, 
NSF-PHY92-00148.



\newpage
\noindent

\begin{figure}
\mbox{
\epsfxsize .45\textwidth \epsfbox{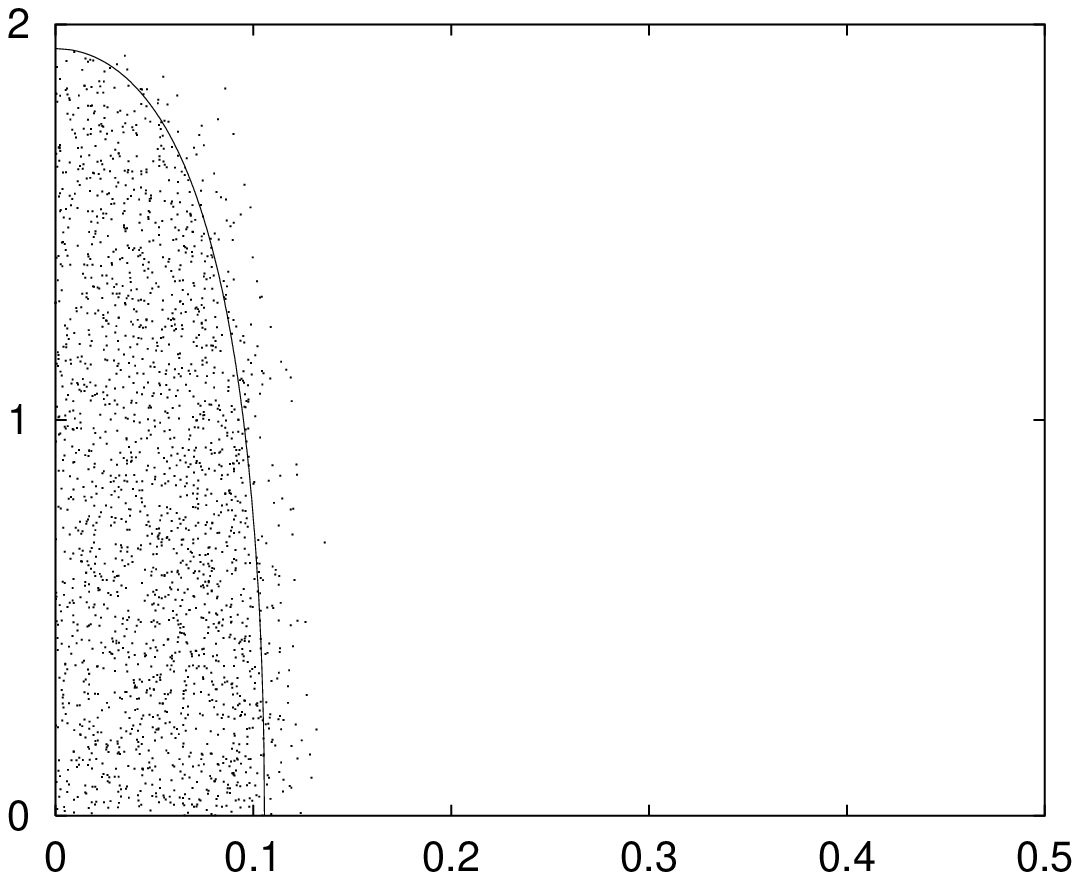}
\epsfxsize .45\textwidth \epsfbox{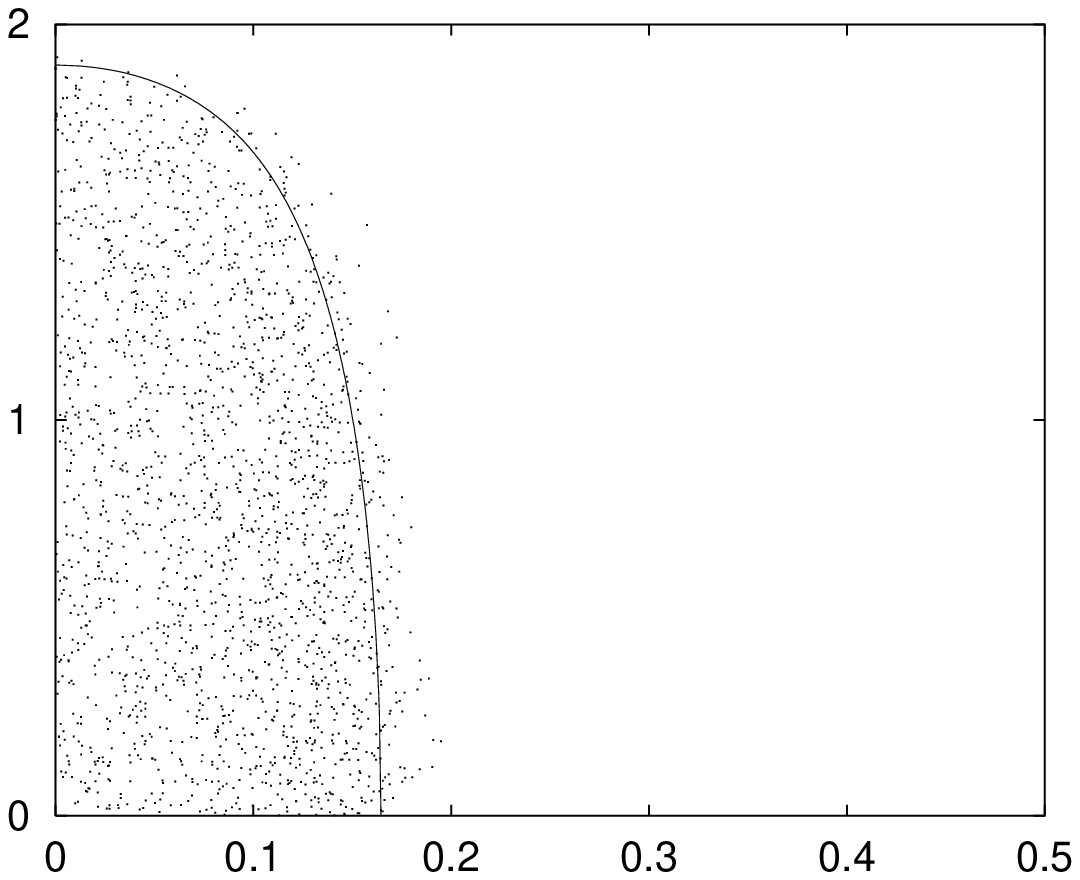}
}\\
\mbox{\hskip .25\textwidth (a) \hskip .45\textwidth (b)}
\vskip 2em
\mbox{
\epsfxsize .45\textwidth \epsfbox{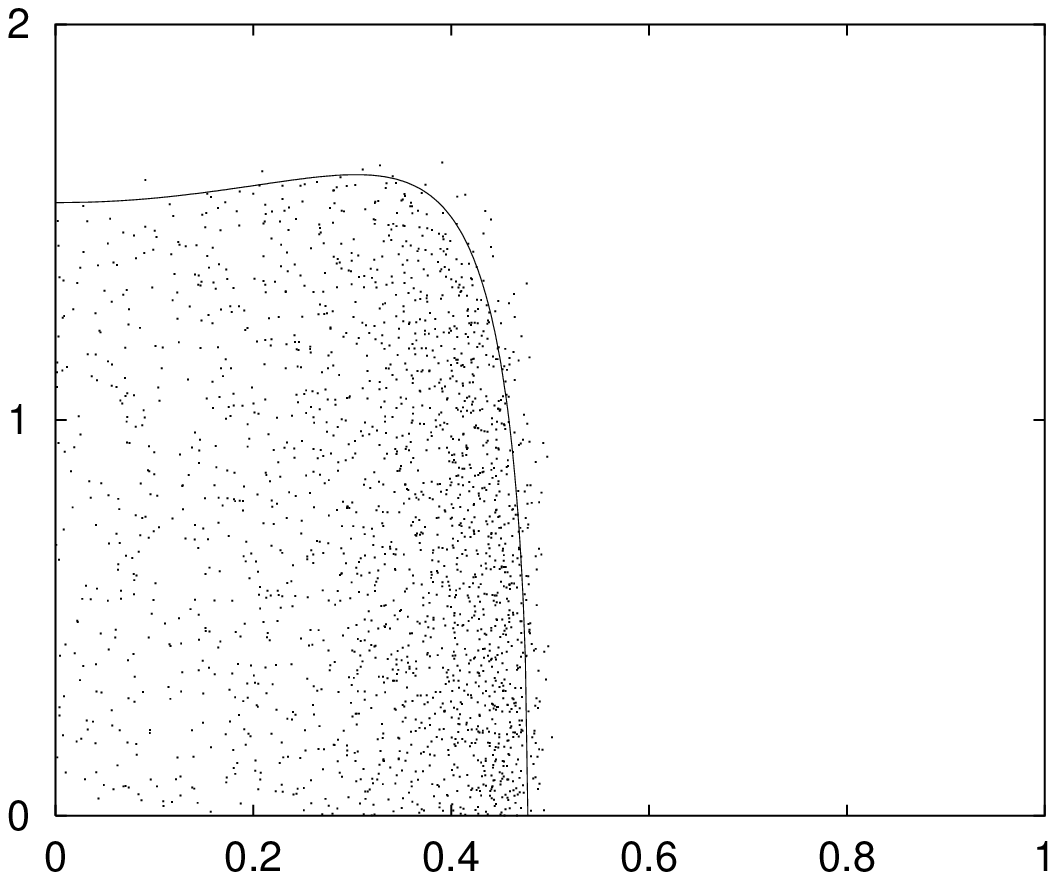}
\epsfxsize .45\textwidth \epsfbox{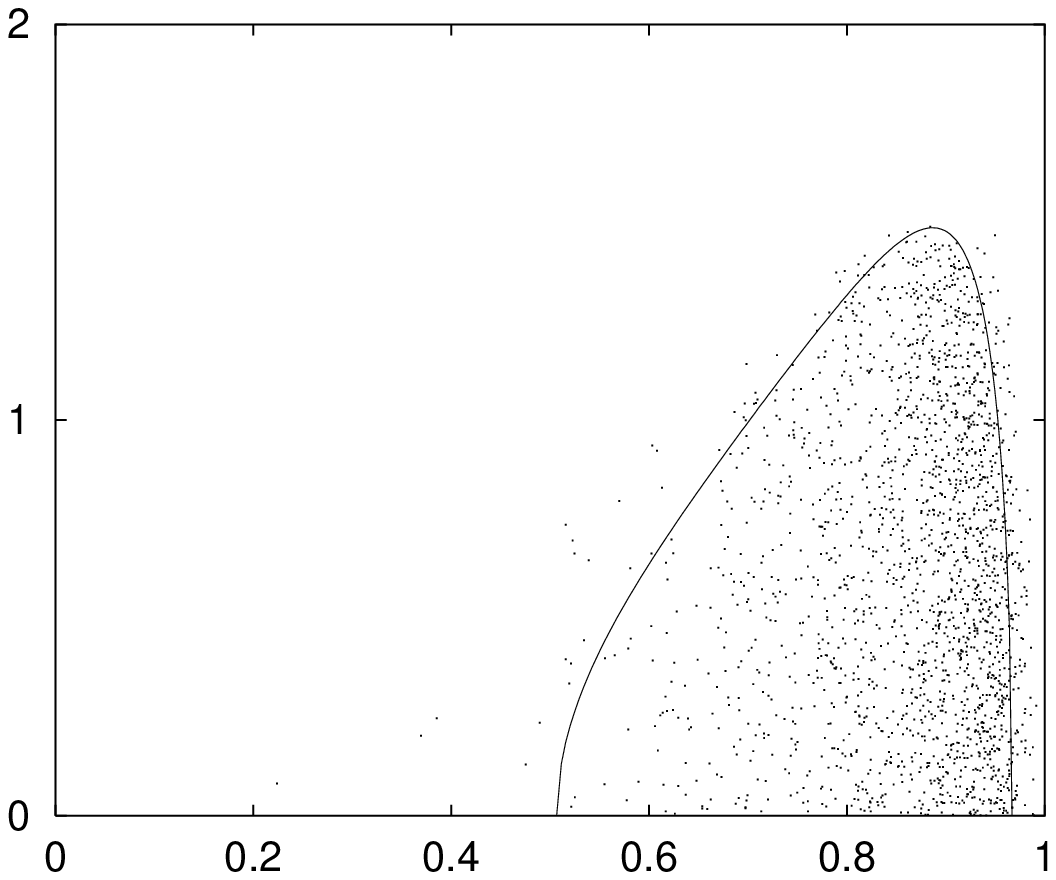}
}\\
\mbox{\hskip .25\textwidth (c) \hskip .45\textwidth (d)}
\vskip 2em
\caption[]{Scatter plots on a complex plane $(x,y)$ of the eigenvalues
of an ensemble of 20 random $100 \times 100$ matrices $D$ at 4 values
of $\mu^2$: 0.06 (a), 0.10 (b), 0.40 (c) and 1.20 (d). The solid curves
follow eq.~(\ref{yx}).}
\end{figure}

\end{document}